\documentclass[12pt, preprint]{aastex}

\usepackage{graphics}
\usepackage{graphicx}
\newcommand{\mgi}{\ion{Mg}{1}}

\shorttitle{Motion of Zodiacal Dust}
\shortauthors{Reynolds et al.}

\begin{document}

\title{New Measurements of the Motion of the Zodiacal Dust}

\author{R. J. Reynolds and G. J. Madsen}
\affil{Department of Astronomy, 475 N. Charter Street,
  University of Wisconsin--Madison, Madison, WI 53706}
\email{reynolds@astro.wisc.edu; madsen@astro.wisc.edu}

\and

\author{S. H. Moseley}
\affil{Code 685, Infrared Astrophysics Branch, Goddard Space Flight Center, 
Greenbelt, MD 20771}
\email{moseley@stars.gsfc.nasa.gov}

\begin{abstract}

Using the Wisconsin H-Alpha Mapper (WHAM), we have measured at high
spectral resolution and high signal-to-noise the profile of the scattered
solar \mgi\ $\lambda$5184 absorption line in the zodiacal light. The
observations were carried out toward 49 directions that sampled the
ecliptic equator from solar elongations of 48$^{\circ}$ (evening sky) to
334$^{\circ}$ (morning sky) plus observations near +47$^{\circ}$ and
+90$^{\circ}$ ecliptic latitude.  The spectra show a clear prograde
kinematic signature that is inconsistent with dust confined to the
ecliptic plane and in circular orbits influenced only by the sun's
gravity.  In particular, the broadened widths of the profiles, together
with large amplitude variations in the centroid velocity with elongation
angle, indicate that a significant population of dust is on eccentric
orbits.
In addition, the wide, flat-bottomed line profile toward the ecliptic pole
indicates a broad distribution of orbital inclinations extending up to
about 30$^{\circ}-$40$^{\circ}$ with respect to the ecliptic plane.  The
absence of pronounced asymmetries in the shape of the profiles limits the
retrograde population to less than 10\% of the prograde population and
also places constraints on the scattering phase function of the particles.  
These results do not show the radial outflow or evening--morning velocity
amplitude asymmetry reported in some earlier investigations. The reduction
of the spectra included the discovery and removal of extremely faint,
unidentified terrestrial emission lines that contaminate and distort the
underlying \mgi\ profile.  This atmospheric emission is too weak to have
been seen in earlier, lower signal-to-noise observations, but it probably
affected the line centroid measurements of previous investigations.

\end{abstract}

\keywords{interplanetary medium---line: profiles---scattering---solar 
system: general}

\section{INTRODUCTION}

The motion of the interplanetary dust particles orbiting the sun contains
clues about their origin and the nongravitational forces affecting their
lifetimes in the solar system, with potential relevance to dust disks
surrounding other stars (e.g., Mann, Gr\"un, \& Wilck 1996; Dermott et al.
1994).  A more accurate understanding of this ``zodiacal cloud'' can also  
impact studies of faint Galactic and extragalactic backgrounds (e.g., 
Haffner et al. 2003; Bernstein, Freedman, \& Madore 2002; Kelsall et al. 
1998).  In theory, Doppler shifts of scattered solar Fraunhofer lines 
in the zodiacal light provide an opportunity to explore and monitor the
large-scale kinematics of interplanetary dust in the inner solar system
through ground-based spectroscopic techniques (e.g., Ingham 1963; James
1969, 1996; Hirchi \& Beard 1987; Clarke et al. 1996), which would be an
important complement to studies of the zodiacal brightness distribution
and the \emph{in situ} detection of interplanetary particles by spacecraft
(Mann 1998; Gr\"un et al. 1997).  In practice, the faint, spatially
extended nature of the zodiacal light, combined with the requirements of
both high signal-to-noise and high spectral resolution to measure the
predicted subtle changes in the position and shape of the line, have made
these observations difficult to carry out.

Early attempts that targeted the H$\beta$ $\lambda$4861 line (e.g., Clarke
et al. 1967; Daehler et al. 1968; Reay \& Ring 1969) encountered the
additional problem that the absorption line was severely contaminated by
emission originating from atomic hydrogen in the earth's atmosphere and
the Galaxy's interstellar medium (Reynolds, Roesler, \& Scherb 1973).  
This led investigators to concentrate their observations on other lines
(James \& Smeethe 1970; Hicks, May, \& Reay 1974; Fried 1978; East \& Reay
1984), especially the \mgi\ line at 5183.62 \AA, which has a large
equivalent width and seemed to have no contaminating atmospheric emission.  
However, except for the conclusion that the motion was primarily prograde,
the resulting information about the orbital properties of the dust was
uncertain and contradictory (see a summary of the situation by Clarke et
al. 1996).

One of the principal difficulties with these earlier observations was the
insufficient throughput of the spectrometers.  This resulted in low
signal-to-noise data and thus prevented precise measurements of line
centroids and shapes, particularly at large solar elongations and at high
ecliptic latitudes, where the zodiacal emission is faintest, but where
observations are crucial for comparisons with predictions of kinematic
models.  Below, we present new, high signal-to-noise observations of the
\mgi\ $\lambda$5184 line in the zodiacal light using the Wisconsin
H$\alpha$ Mapper (WHAM) Fabry-Perot facility at the National Optical
Astronomy Observatory on Kitt Peak in Arizona.  WHAM's unprecedented
combination of high throughput and high spectral resolution reveal clearly
resolved \mgi\ line profiles that are shifted and broadened by the motion
of the zodiacal dust.

\section{THE WISCONSIN H$\alpha$ MAPPER (WHAM) OBSERVATIONS}

WHAM consists of a 15 cm aperture, dual-etalon Fabry-Perot spectrometer
coupled to a high efficiency CCD camera and fed by a dedicated 0.6 m
telescope.  The dual-etalon design, when compared to traditional single
etalon Fabry-Perot systems, provides superior ghost and order rejection,
extends the free spectral range, and suppresses the Lorentzian-like wings
of the spectral response function.  The CCD is used as a multichannel
detector to record a 12 km s$^{-1}$ (0.2 \AA\ at 5184 \AA) resolution
spectrum over a 200 km s$^{-1}$ (3.5 \AA) spectral window that can be
centered on any wavelength between 4800 \AA\ and 7300 \AA.  The 15 cm
diameter etalons provide maximum throughput, which when combined with the
0.6-m aperture telescope, results in a 1\fdg0 diameter beam on the
sky.

Although built and operated primarily for studies of faint diffuse
emission lines, particularly H$\alpha$ from the Galaxy's interstellar
medium, WHAM is ideally suited for high spectral resolution observations
of any faint spatially extended source in the night sky, including comets
(Morgenthaler et al. 2001), the earth's atmosphere (Nossal et al. 2001),
and the zodiacal light.  The facility is completely remotely operated, and
all the observations presented here were obtained with WHAM controlled
from Madison, Wisconsin.  Additional information about WHAM can be found
in Haffner et al. (2003) and Tufte (1997).

Spectra centered on \mgi\ $\lambda$5184 were obtained on the two moonless
nights of 2002 November 4 and 2002 November 5.  The spectra sample 49
directions near the ecliptic equator between elongations $\epsilon$ =
47\fdg5 (evening sky) and $\epsilon$ = 333\fdg9 (26\fdg1 west of the sun
in the morning sky).  Two directions at high ecliptic latitude were also
observed, $\epsilon \approx$ 272$^{\circ}$, $\beta \approx$ +47$^{\circ}$
and the north ecliptic pole (NEP).  In addition, observations of the
relatively bright twilight sky were obtained to provide a calibration
spectrum close to representing the
unperturbed solar line profile.  The directions of the zodiacal
observations and the integration time for each spectrum are listed in the
first three columns of Table 1.  A 30$^{\circ}$ gap in the data near
elongation 225$^{\circ}$ is due to interference by stars near the plane of
the Milky Way, which produced anomalies in the spectra.  The Galactic
plane intersected the ecliptic plane at $\epsilon = 50^{\circ}$ and
$\epsilon = 230^{\circ}$ at Galactic longitudes 6$^{\circ}$ and
186$^{\circ}$, respectively.

A twilight spectrum is shown in Figure~1. The absorption feature is
dominated by the \mgi\ line at 5183.62 \AA, with weak contributions from
\ion{Fe}{1}\ at 5184.27 \AA\ and \ion{Cr}{1}\ at 5184.59 \AA,
respectively.  A potential problem in using twilight spectra to represent
the unperturbed source spectrum is the contamination of the daylight and
twilight Fraunhofer lines by a pseudo-continuum emission from inelastic
scattering (i.e., rotational Raman scattering) of sunlight by atmospheric
nitrogen and oxygen molecules. Both observations and theory suggest that
this so-called ``Ring effect'' tends to fill in the Fraunhofer lines at
about the 10\% level with minimal change to the width of the line (e.g.,
Grainger \& Ring 1962; Pallamraju, Baumgardner, \& Chakrabarti 2000;
Sioris \& Evans 1999).  To investigate this we compare our twilight
spectrum in Figure 1 with a high resolution unscattered solar spectrum
convolved with WHAM's $12~$km~s$^{-1}$ wide spectral response profile.  
After scaling to match the ``continua'' and the profile minima within the
observed spectral interval, we found the profiles to be nearly coincident
(see Fig. 1).  Thus both the twilight profile and the convolved direct
solar profile appear to be good representations of the unperturbed solar
spectrum and useful for comparison with the Doppler perturbed zodiacal
profiles.  The Ring effect, a phenomenon affecting the spectrum of
sunlight that has been scattered by atmospheric molecules, is not relevant
to the zodiacal light, which is scattered by interplanetary dust
particles.  Zero on the abscissa in Figure 1 and in all following spectra
is defined by the centroid of the core of the \mgi\ component in the
twilight spectra, with a small ($\sim 0.2~$km~s$^{-1}$) correction for the
observed Doppler shifts associated with the earth's rotation.  The
uncertainty in this calibration is less than $\pm$0.02 \AA\
($\pm$1~km~s$^{-1}$).

\section{THE ZODIACAL SPECTRA}

\subsection{Contamination by Weak Atmospheric Emission Lines}

The average spectrum east of the sun near $\epsilon \approx 100^{\circ}$
and the average spectrum at the corresponding elongation west of the sun
near $\epsilon \approx 260^{\circ}$ are shown superposed in Figure 2a.  
The prograde kinematic signature of the zodiacal dust is clearly seen,
with the profile in the evening spectrum (filled triangles) red shifted
and the profile in the morning spectrum (filled circles) blue shifted with
respect to the centroid of the unperturbed solar \mgi\ line.  
However, in addition to being shifted (and broadened), these spectra
exhibit a number of extremely weak, relatively narrow emission lines.  
The fact that these emission features are present at the same wavelengths
in both of the spectra in Figure 2a implies that they are not associated
with the Doppler shifted light from the zodiacal dust and instead
originate from the earth's atmosphere.  Previous studies of interstellar
emission with WHAM have shown that such weak (intensities $\lesssim 0.1$
R\footnote{1 R = $10^6$/4$\pi$ photons s$^{-1}$ cm$^{-2}$ sr$^{-1}$}),
unidentified atmospheric emission lines are present throughout the visible
spectrum (e.g., Hausen et al. 2002;  Madsen et al. 2004, in prep). They
distort the shapes of observed zodiacal and interstellar line profiles,
modify their measured intensities and equivalent widths, and in
observations with insufficient spectral resolution produce a
pseudo-continuum of order $10^{-8}$ erg cm$^{-2}$ s$^{-1}$ sr$^{-1}$
\AA$^{-1}$.  Therefore, before proceeding with the analysis of the
zodiacal line profiles, it was necessary to remove this contaminating
terrestrial emission from each spectrum.

To characterize more accurately the spectrum of this foreground emission,
all spectra in directions with the faintest zodiacal emission (i.e., $
88^{\circ} < \epsilon < 272^{\circ}$) were averaged together, producing
the very high signal-to-noise composite spectrum in Figure 2b, where all
the zodiacal absorption features, including the Fe and Cr lines, are
highly Doppler broadened into a single, smooth profile.  From this
spectrum it was possible to identify the narrow terrestrial emission
features against the underlying, Doppler smeared zodiacal spectrum by
fitting a smooth, single-gaussian profile to the spectrum in Figure 2b in
such a way that the residuals were near zero or positive at all
velocities.  The residuals, indicated by the solid line in Figure 2c, then
represent the spectrum of the foreground terrestrial emission.  The
sharpness of the features in Figure 2c provides additional confirmation
that they are not associated with the highly broadened zodiacal profile.  
Also plotted in Figure 2c are the maximum and minimum allowed terrestrial
emission spectra that are consistent with the spectrum in Figure 2b.

Although too weak to have been identified in earlier, lower 
signal-to-noise observations, this terrestrial emission, if not removed, would
clearly distort the apparent centroid and shape of the underlying zodiacal
absorption feature (see Fig. 4a below). Therefore, to obtain
uncontaminated zodiacal profiles, we subtracted the best fitted
terrestrial emission spectrum (solid line in Fig. 2c) from each of the
spectra listed in Table 1.  This process treated each spectrum the same
and did not take into account possible temporal variations in the
intensity of this emission or variations in the relative strengths of the
individual emission components.  However, because the emission is
extremely weak, there was no way to identify and remove it from each
spectrum independently.  When the observations were separated by halves of
the night, by day, and by high and low ecliptic latitude, the ``best fit''
spectrum in Figure 2c provided good removal of the emission features
within the uncertainties, suggesting that this spectrum fairly
approximates the terrestrial contamination in each individual spectrum.

\subsection{Pure Zodiacal Spectra}

Figure 3 shows a sample spectrum after removal of the contaminating
atmospheric emission as described above.  The abscissa is in units of
radial velocity, where zero corresponds to the centroid of the unperturbed
\mgi\ $\lambda$5184 component in the solar spectrum (Fig. 1).  To obtain
the centroid velocity (V), full width at half maximum (W), and area (A) of
the zodiacal \mgi\ absorption component, a least squares three-gaussian
fit, representing the \mgi\ and the two weak Fe~I and Cr~I features, was
made to the spectrum.  During the fit the relative centroid positions and
the relative areas of the three components were held fixed at the values
found in the twilight spectrum, and the widths of the two minor
Fe~I and Cr~I features were held fixed at 55 km s$^{-1}$, the approximate
mean broadening of the zodiacal spectra.  The fitting procedure also took
into account the 12 km s$^{-1}$ wide instrumental response function of the
WHAM spectrometer and a small correction for Doppler shifts due to the
earth's rotation.  The resulting best fit (solid curve) and residuals are
included in Figure 3.  Values for the best fit parameters of the \mgi\
component for this and all the other spectra are listed in Table 1.

The resulting radial velocities, widths, and areas of the \mgi\ component
along the ecliptic equator are plotted in Figures 4a, 4b, and 4c,
respectively.  The uncertainties associated with these results are the
combined uncertainty associated with random errors due to Poisson noise,
the placement of the continuum, plus the systematic error associated with
the subtraction of the atmospheric emission.  In this last case the error
was estimated by fitting each spectrum two more times, after subtracting
the ``minimal'' and the ``maximal'' versions of the atmospheric emission
spectrum (Fig. 2c).  We found that the uncertainty in the radial velocity
is dominated by the systematic error associated with the removal of the
atmospheric emission and that errors in the widths and areas are dominated
by the uncertainty in the placement of the continuum.  The uncertainties
listed in Table 1 and plotted in Figures 4a and 4b below indicate only the
systematic error due to the removal of the atmospheric emission.  In most
observations, the subtraction of the ``minimal'' (``maximal'') terrestrial
spectrum instead of the ``best'' terrestrial spectrum results in
systematically less (more) positive centroid velocities, narrower (wider)
line widths, and slightly larger (smaller) areas.  Also plotted are the
resulting velocities (open squares in Fig.  4a) when \emph{no} correction
was made for the atmospheric emission, to illustrate the significant
effect this emission can have on the derived results if it is not removed
from the spectra. The effect on W and A is less dramatic, due in part to
the fact that, while some of the emission lines fill in the zodiacal
absorption profile slightly, other lines raise the ``continuum''.  The
scatter in the values for the line widths and areas, which is
significantly larger than the plotted (systematic) error, reflects
primarily the uncertainties in W and A associated with continuum
placement.

Figure 4a reveals a well-characterized velocity vs. elongation kinematic
signature for the line centroids indicative of prograde orbits, and Figure
4b shows significant broadening of the line profiles with respect to the
unperturbed solar spectrum at all elongations, including 180$^{\circ}$.  
The radial velocities pass through zero at $\epsilon \approx 54^{\circ}$,
near 180$^{\circ}$, and at $\epsilon \approx 315^{\circ}$, indicating a
small (10$^{\circ}$) evening--morning asymmetry in the elongation at which
the radial velocity passes through zero.  These data show no evidence for
a general outflow (i.e., a mean positive velocity shift at 180$^{\circ}$)  
or an evening--morning asymmetry in the velocity amplitudes as reported in
some earlier studies (East \& Reay 1984; Hicks et al. 1974).  Within the
uncertainty associated with the removal of the atmospheric emission (see
above), the radial velocity at 180$^{\circ}$ is consistent with zero and
the maximum amplitudes of the radial velocities are anti-symmetric,
reaching approximately +13 km s$^{-1}$ at $\epsilon \approx 100^{\circ}$
and approximately $-12$ km s$^{-1}$ at $\epsilon \approx 260^{\circ}$. The
area of the \mgi\ absorption profile is a tracer of the zodiacal continuum
intensity, and Figure 4c shows evidence for the Gegenschein and the steep
rise in zodiacal intensity for directions within about 70$^{\circ}$ of the
sun.  The absolute intensity calibration of these observations is
accurate to only about 20\%, and zero intensity relative to the
continuum is uncertain due to an added flat continuum from parasitic
(out of band) light. Because the intensity distribution of the
zodiacal light has been well studied by others, the remainder of this
paper focuses on some of the implications of the new kinematic information
revealed by the radial velocities and line widths of the \mgi\ profile
presented in Figures 4a and 4b.

\section{DISCUSSION}

\subsection{Orbital Eccentricities}

The orbital properties of the zodiacal dust, including the
eccentricity, inclination, and the effect of radiation pressure, can
be explored by comparing these results with line profiles and Doppler
shifts calculated from models of the distribution, kinematics, and
scattering properties of the dust (e.g., James 1969; Bandermann \&
Wolstencroft 1969; Rodriguez \& Magro 1978; Hirschi 1985; Hirschi \&
Beard 1987; Mukai \& Mann 1993;  Clarke et al. 1996).  In Figure 5 the
observed velocity centroids V and line widths W are compared with
models by Hirschi \& Beard (1987) and Hirschi (1985), who have made
the most detailed predictions of these two parameters.  They
considered a variety of models with isotropically scattering particles
in prograde orbits confined to the ecliptic plane.  We consider four
of their models here: a) particles on circular orbits and influenced
only by solar gravity, b) particles on elliptical orbits having a
uniform distribution of eccentricities between 0 and 1, c) particles
on elliptical orbits with a distribution of eccentricities weighted
toward low eccentricities, and d) particles on circular orbits but
influenced by radiation pressure, that is, feeling a lower effective
gravity.  These models are represented by the solid, dotted,
dot-dashed, and dashed curves, respectively, in Figure 5.

The data clearly rule out pure circular orbits (model a; solid line).  
The predicted amplitude of the radial velocity variations is too small,
the elongations at which the predicted radial velocity curve passes
thorough zero are too far from the sun, and the widths of the predicted
profiles are too narrow to fit the observations.  Particles in circular
orbits, but moving at sub-Keplerian speeds due to radiation pressure
(model d; dashed line), would exhibit a larger radial velocity amplitude
and intercept zero velocity closer to the sun (Fig. 5a), depending on the
magnitude of the effect.  However, the predicted line widths from such
circularly orbiting dust would still remain far too narrow to fit the
observations (Fig. 5b).

These results agree best with models in which the zodiacal dust is moving
on elliptical orbits.  This conclusion is supported most convincingly by
the observed broadening of the line profiles.  As pointed out by Hirschi
\& Beard (1987), it is the line widths, especially near elongation
180$^{\circ}$, that clearly distinguish spectra of elliptically orbiting
dust from spectra of dust in circular orbits, including dust in circular
orbits experiencing radiation pressure.  Figures 4b and 5 and Table 1 show
that the profiles are broadened significantly with respect to the
unperturbed solar profile and with respect to profiles predicted by models
in which the dust is only in circular orbits.  For example, Figure 6
compares the average spectrum for the five observations within 8$^{\circ}$
of $\epsilon$ = 180$^{\circ}$ with the unperturbed solar spectrum.  If all
the orbits were circular and centered on the sun, the profiles would be
nearly identical.  The broadening of the zodiacal profiles can only be
explained by particles having radial components to their orbital motion.

\subsection{Radiation Pressure and Orbital Inclinations}

Orbital eccentricity alone is not sufficient to account for
these observations.  Although the predicted radial velocity variation
from a dust population with a uniform distribution of orbital
eccentricities (model b; dotted line) matches the observed radial
velocities very well (Fig. 5a), the line widths predicted by this
model are too large, especially near $180^{\circ}$ (see Fig. 5b).  
The model that is weighted toward lower eccentricities (model c;
dot-dashed) provides a much better fit to the observed widths overall;
however, that model does not predict a velocity amplitude as large as
is observed.  Models that combine eccentric orbits (to account for the
line width) with an influence from radiation pressure (to increase the
velocity amplitude) appear to be needed to obtain the best fit to the
observed velocity variations and the line widths simultaneously (see
Table 3, column 2 in Hirschi \& Beard 1987).  Another orbital
parameter, which these models did not include, is inclination.  
Particles with a distribution of orbital inclinations will have
smaller projected velocities along sightlines in the ecliptic plane.
This will also increase the radial velocity amplitude (Rodriguez \&
Magro 1978) and thus mimic the effect of radiation pressure in
observations confined to the ecliptic equator. Perhaps a combination
of eccentricities and inclinations may even obviate the need for a
significant effect from radiation pressure.  The observations at high
ecliptic latitude are therefore important for both probing the
inclinations of the orbits and quantifying the influence of radiation
pressure.  Unfortunately, there are not yet models incorporating both
eccentricity and inclination that predict both W and V.

A wide distribution of orbital inclinations is implied by the broad line
profiles toward $\epsilon \approx 272^{\circ}$, $\beta \approx$
$47^{\circ}$ and the NEP (Table 1).  The profile toward the NEP not only
has one of the largest widths, but unlike profiles in all the other
directions, has a distinctly flat-bottom shape.  The NEP spectrum and the
unperturbed solar spectrum are shown in Figure 7.  This substantial
broadening of the profile toward the ecliptic pole, greater than that
toward the antisolar direction, implies a population of particles that
have significant components of their orbital velocities projected
perpendicular to the ecliptic plane.  For an orbital speed of 30 km
s$^{-1}$ near 1 AU, and neglecting the radial motion associated with
eccentricity, the approximately $\pm 15 - 20$ km s$^{-1}$ broadening at
the base of the NEP profile suggests a distribution of inclinations
extending up to about 30$^{\circ} - 40^{\circ}$ with respect to the
ecliptic plane.  This is close to, but somewhat broader than the
distribution derived from the Galileo and Ulysses \emph{in situ}
interplanetary dust impact detectors (Gr\"un et al. 1997).

\subsection{Line Profile Asymmetries}

After removal of the atmospheric emission and taking into account the weak
solar Fe and Cr features on the red side of the \mgi\ line (Fig. 1), we see
no pronounced asymmetries in the shape of the zodiacal profiles.  This
places additional constraints on the properties of the zodiacal dust.  
For example, the nature of the scattering phase function of the dust is
uncertain (e.g., Hahn et al. 2002), and Clarke et al. (1996) have shown that
significant profile asymmetries will be present if the scattering function
derived by Hong (1985) is used instead of isotropic scattering.  In
particular, they predicted that profiles between $\epsilon \approx 50^{\rm
o} - 70^{\circ}$ and in the corresponding interval at $\epsilon \approx
290^{\circ} - 310^{\circ}$ would have clearly extended wings on the red
and blue sides, respectively.  Another source of profile asymmetry would
be a population of dust in retrograde motion.  James (1969) calculated the
radial velocities that would be associated with particles on retrograde
orbits and showed that they would produce a Doppler shifted absorption
feature centered near $+55$ km s$^{-1}$ and $-55$ km s$^{-1}$ at $\epsilon \approx
50^{\circ} - 70^{\circ}$ and $\epsilon \approx 290^{\circ} - 310^{\circ}$,
respectively.

To search for these asymmetries, the mean spectrum within each of these
two elongation intervals is plotted in Figure 8.  To make the comparison
easier the mean spectrum toward $\epsilon \approx 50^{\circ} - 70^{\circ}$
was shifted 9.0 km~s$^{-1}$ to the blue so that the minima of the two
profiles coincide.  Within the scatter of the data points, the profile
shapes appear to be nearly identical, indicating that there are no
pronounced asymmetries produced by retrograde particles or the scattering
discussed above.  For example, the absence of a significant depression on
the red side of the evening spectrum (open circles) relative to that part
of the morning spectrum (filled circles), and \emph{vice versa} for the
blue side, indicate that any population of retrograde particles must be
less than 10\% the prograde population.

The spectra toward 180$^{\circ}$ and the NEP show marginal evidence for
extra absorption on the red side of the profiles near +30 km s$^{-1}$
(Figs. 6 and 7).  This could be an indication of a small population of
particles with a radial outflow (Hirschi 1985); however, given current
uncertainties associated with the removal of terrestrial emission (e.g.,
Fig. 6), it is difficult to judge its significance.  A more extensive
observational program, including studies of more than one Fraunhofer line,
will be needed to characterize more accurately such subtle asymmetries
(see below).

\section{CONCLUSIONS AND FUTURE OBSERVATIONS}

Observations of the \mgi\ $\lambda$5184 line with WHAM have demonstrated
that high throughput spectroscopic measurements of scattered solar
Fraunhofer line profiles in the zodiacal light can be a useful tool for
exploring the properties of the zodiacal dust, as suggested by Ingham
(1963), James (1969) and others.  These observations, combined with models
of the zodiacal cloud, have probed or placed constraints on the retrograde
and prograde populations, the orbital eccentricities and inclinations, the
effects of non-gravitational forces, and the scattering properties of the
particles.  Furthermore, this characterization of the Doppler distortions
of Fraunhofer profiles in the zodiacal light should facilitate a more
accurate removal of zodiacal contamination in other astrophysical
observations (e.g., see Haffner et al. 2003).  The detection of the
extremely weak telluric emission lines also implies that caution must be
used in the interpretation of line profiles, equivalent widths, line
intensities, and the nature of the ``continuum'' in low resolution spectra
of zodiacal light and other low surface brightness astrophysical sources.

These results also point to ways in which observations can be further
improved.  For example, integration times can be significantly lengthened
to increase the signal-to-noise, and the bandwidth of the spectra could be
expanded to make it possible to define the continuum more accurately.  A
fuller sampling at higher ecliptic latitudes would also help to
characterize more accurately particle orbits inclined to the ecliptic
plane (Rodriguez \& Magro 1978).  Systematic errors associated with the
removal of the atmospheric emission lines, the largest source of
uncertainty in the measurements of radial velocities and profile shapes,
could be reduced by observing two Fraunhofer lines (e.g., \mgi\
$\lambda$5173 in addition to \mgi\ $\lambda$5184).  Both lines are
modified identically by scattering from the zodiacal dust but are affected
differently by the atmospheric foreground emission.  A complete set of
such spectra obtained multiple times over the course of a year or more
would also allow a search for more subtle effects associated with possible
azimuthal asymmetries in the zodiacal cloud, the inclination of the
earth's orbit with respect to the dust plane, and the eccentricity of the
earth's orbit (e.g., Dermott et al. 1994; Kelsall et al. 1998). In
addition, because observations toward the same elongation at different
times of the year sample different parts of the Milky Way, observations
spread throughout the year could help to characterize and remove any
spectral features associated with the background starlight (Mattila 1980).
A more accurate unperturbed source spectrum (free of any ``Ring effect'')
could be obtained from observations of the moon, rather than the twilight
sky.  The WHAM facility provides an opportunity to carry out such an
investigation.  The proper interpretation of these data also requires a
more complex grid of models, incorporating a range of orbital and
scattering parameters and particle distributions (e.g., Haug 1958) that
predict the velocities and shapes of the zodiacal line profiles over the
sky.

\section{ACKNOWLEDGMENTS}

We thank Matt Haffner, Kurt Jaehnig, and Steve Tufte for their valuable
contributions to the operation of WHAM, Alexander Kutyrev for his 
assistance and encouragement, and an anonymous referee for helpful 
comments.  We are also grateful to Dean Hirschi 
for providing a copy of his PhD thesis. The WHAM program is supported 
by the National Science Foundation through grant AST-0204973.

\clearpage

\begin{figure}[t]
\epsscale{0.75}
\plotone{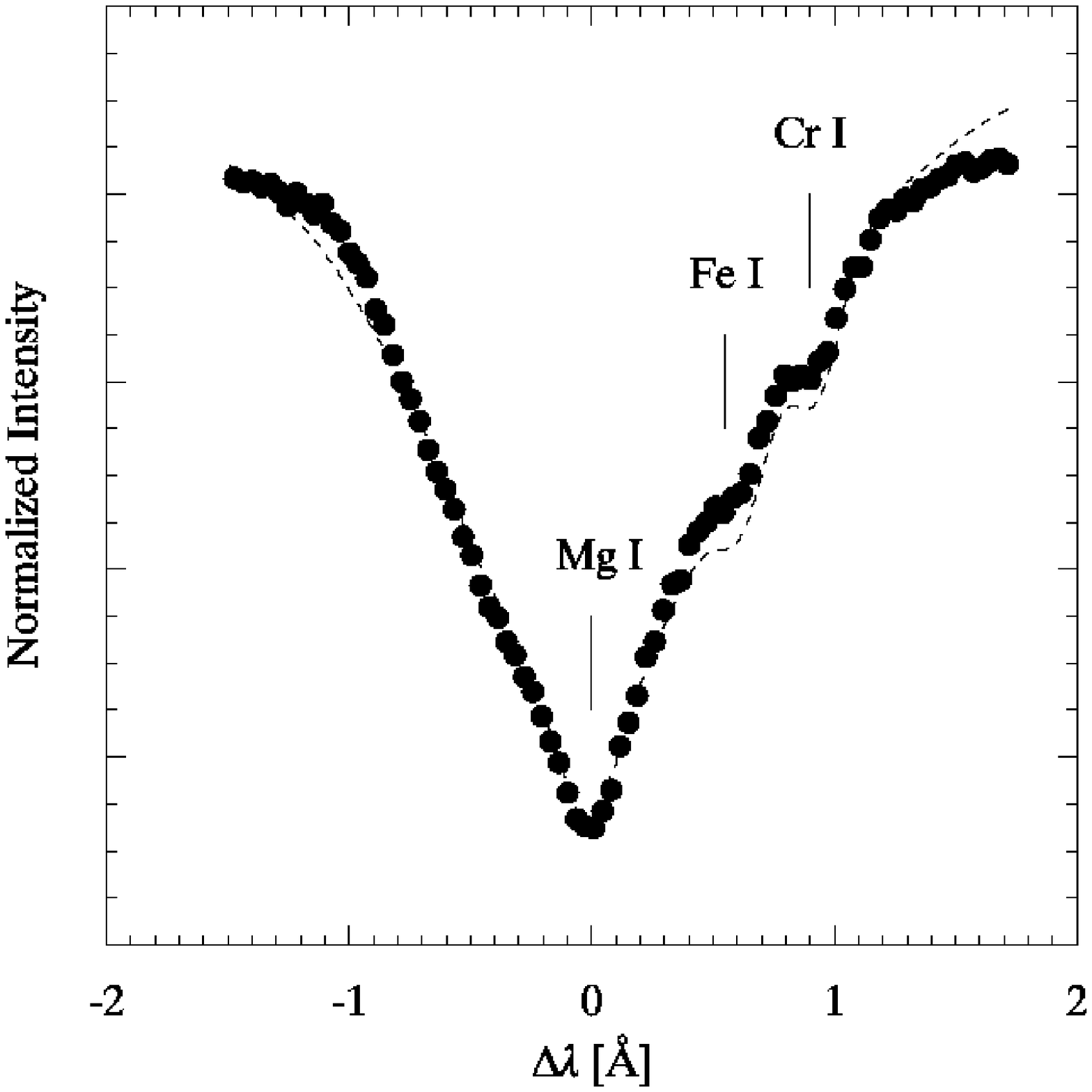}
\caption{Unperturbed solar spectrum from the bright twilight sky (filled 
circles), 
showing the \mgi\ profile plus the two weaker, narrower lines from \ion{Fe}{1} and 
\ion{Cr}{1}.  For comparison, the dashed curve shows the direct, high 
resolution solar spectrum from Delbouille, Neven, \& Roland (1973) 
convolved with WHAM's spectral response function (see \S2).} \end{figure}

\clearpage

\begin{figure}[t]
\epsscale{0.75}
\plotone{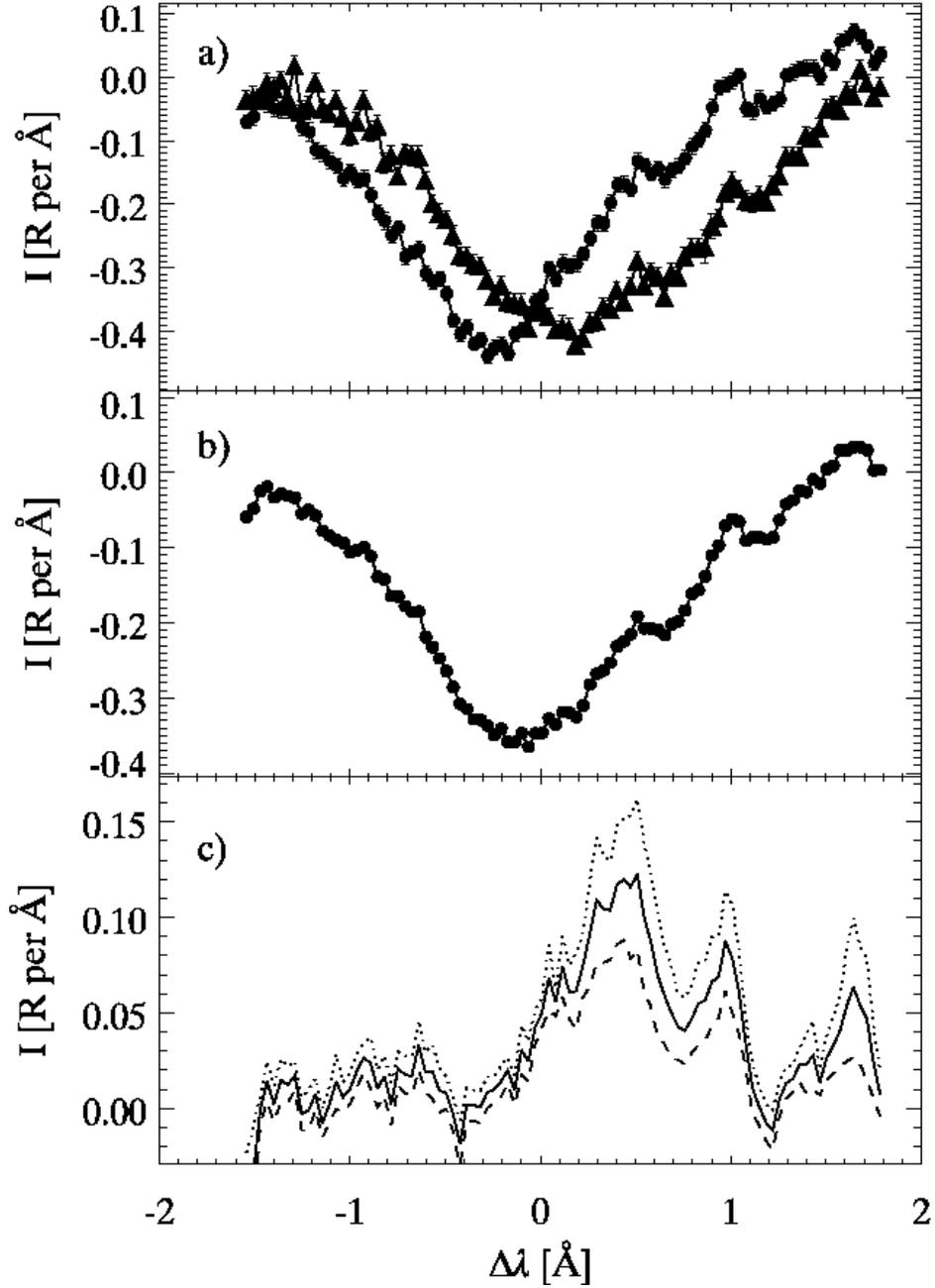}
\caption{a) Average spectrum between $\epsilon \approx 88^{\circ}-108^{\circ}$
(circles) and the average spectrum between $\epsilon \approx 252^{\circ}-264^{\circ}$
(triangles), showing weak, narrow, unshifted terrestrial emission lines
superposed on the scattered solar profile, broadened and shifted by the
motion of the zodiacal dust.  b) Average of all spectra beween
elongations 88$^{\circ}$ and 264$^{\circ}$, which shows more clearly 
the
contaminating terrestrial emission lines superposed on the kinematically
smeared, high signal-to-noise zodiacal spectrum.  c) ``Best fit''
(solid line) atmospheric emission spectrum extracted from the 
spectrum in Fig. 2b (see
text).  The dotted line represents the maximal atmospheric contamination 
and the dashed line the minimal atmospheric contamination consistent with 
the spectrum in Fig. 2b.  1~R per \AA\ corresponds to $3 \times 
10^{-7}$ 
erg cm$^{-2}$ s$^{-1}$ sr$^{-1}$ \AA$^{-1}$ at 5184 \AA.
} 
\end{figure}

\clearpage

\begin{figure}[t]
\epsscale{0.75}
\plotone{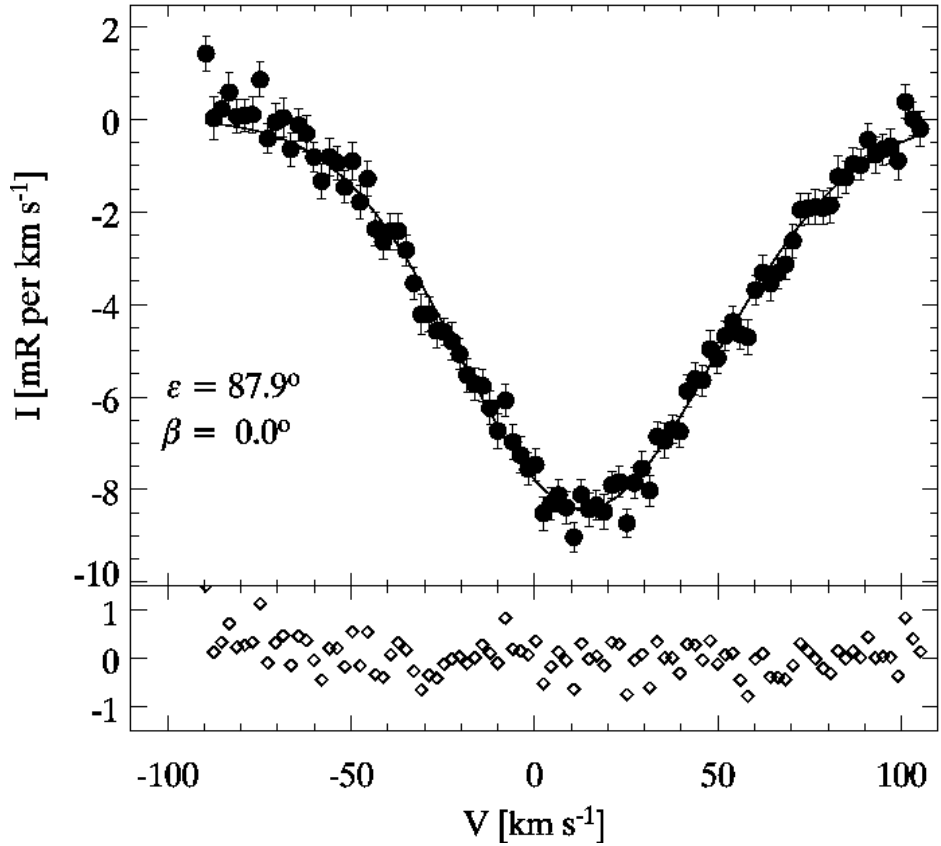}
\caption{Sample spectrum of the zodiacal light toward $\epsilon \approx
88^{\circ}$ with the terrestrial emission removed.  The integration 
time was 600~s.  The best
three-gaussian (\mgi, \ion{Fe}{1}, and \ion{Cr}{1}) fit (solid line) and the resulting
residuals are also indicated. The error bars represent the
$\pm 1 \sigma$ Poisson noise.
}
\end{figure}

\clearpage

\begin{figure}[t]
\epsscale{0.75}
\plotone{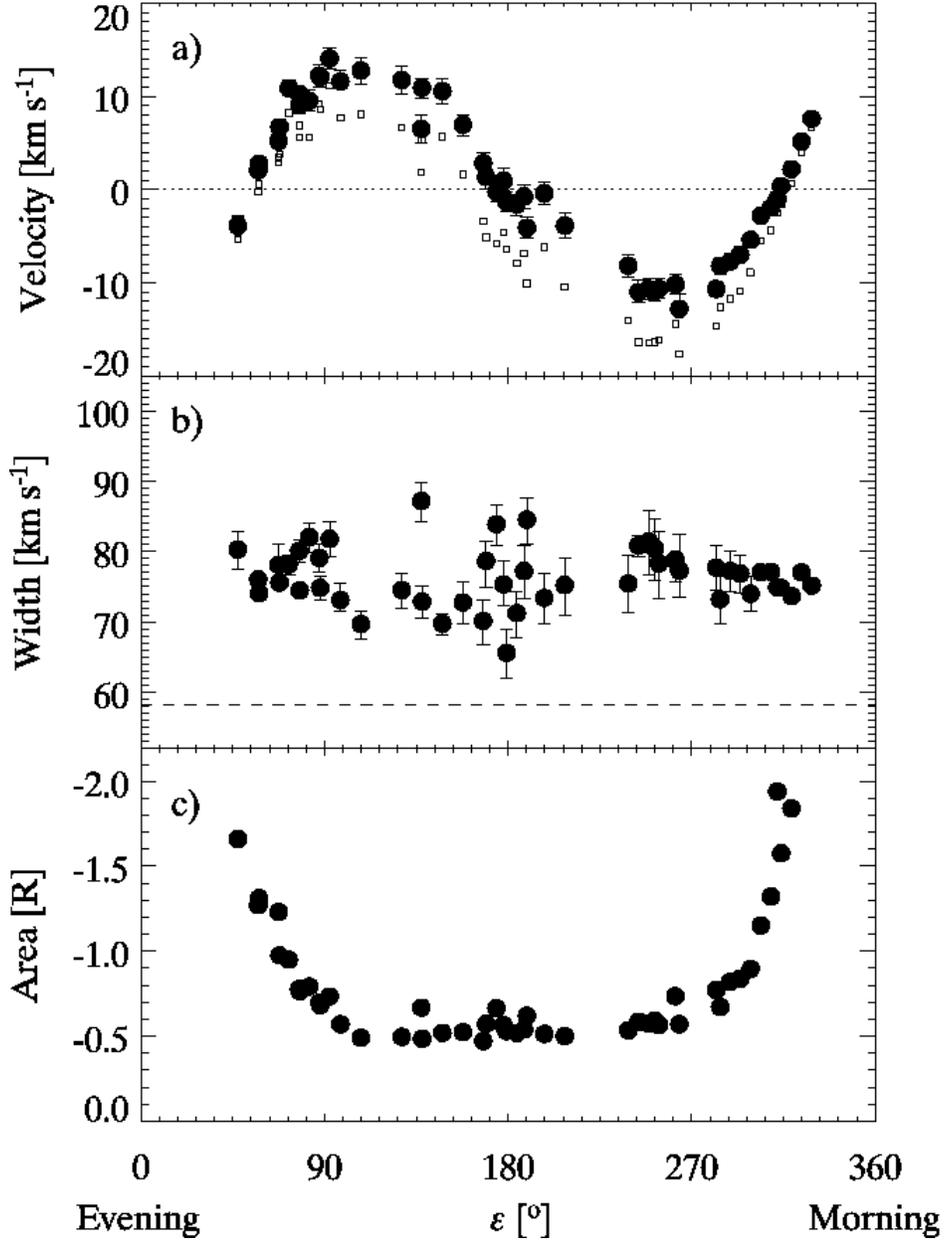}
\caption{Best fit centroid radial velocity (V), full width at half
maximum (W), and area (A) for the \mgi\ line are plotted for each of the
zodiacal spectra near the ecliptic equator.  The error bars represent the
systematic uncertainty listed in Table 1 associated with the removal 
of the 
contaminating
atmospheric emission (see \S3).  The open squares in Fig. 4a show 
the resulting velocities if the atmospheric emission is not 
removed.  The dashed line in Fig. 4b represents the width of the \mgi\ 
component in the unperturbed solar spectrum.
}
\end{figure}

\clearpage

\begin{figure}[t]
\epsscale{0.5}
\plotone{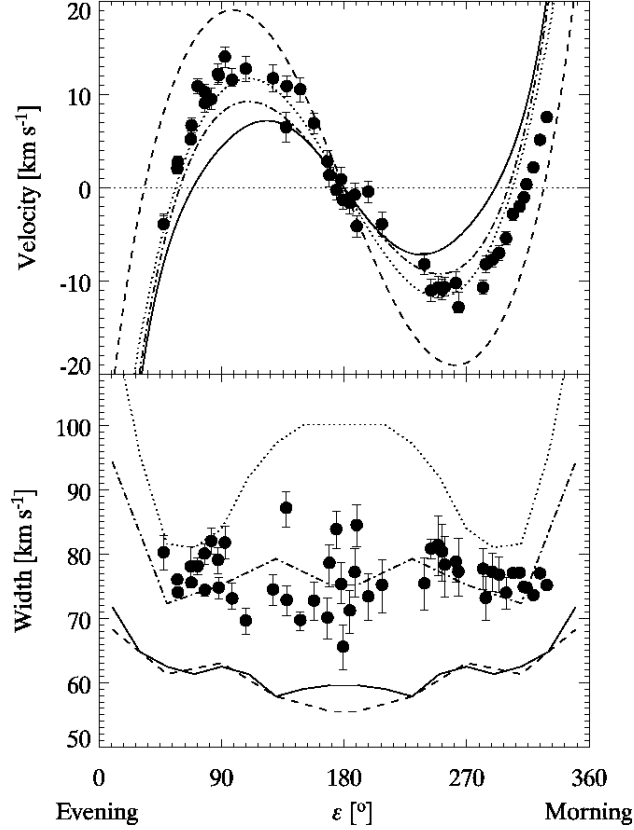}
\caption{Predictions for the velocities and widths of the \mgi\ line from
models of the zodiacal dust by Hirschi \& Beard (1987) and Hirschi (1985):  
\emph{solid line} (model a): particles on circular orbits and influenced
only by solar gravity; 
\emph{dotted line} (model b): particles on elliptical orbits
with a uniform distribution of eccentricities between 0 and 1, randomly
distributed perihelions and no radiation pressure; \emph{dot-dashed line}
(model c):  similar to model b above, but weighted toward lower
eccentricities such that n(e) $\sim$ 1/e; \emph{dashed line} (model d): 
particles on circular
orbits with a size distribution proportional to s$^{-4}$ down to about
0.5 $\mu$m 
and density 1 g cm$^{-3}$, resulting in a significant influence by solar radiation
pressure that reduces the effective gravity.
In all models the particle
motion is parallel to the ecliptic plane, and in models a, b, and c, the
particle number density decreases with distance from the sun as n(r)
$\sim$ r$^{-1.5}$; taken from Table 2 in Hirschi \& Beard (1987).  For
model d, n(r) $\sim$ r$^{-2}$, and the velocities are taken from Table 4
in Hirschi (1985). A seventh order polynomial passing through all the 
model data points was used to interpolate between model values for 
velocities.
}
\end{figure}

\clearpage

\begin{figure}[t]
\epsscale{0.75}
\plotone{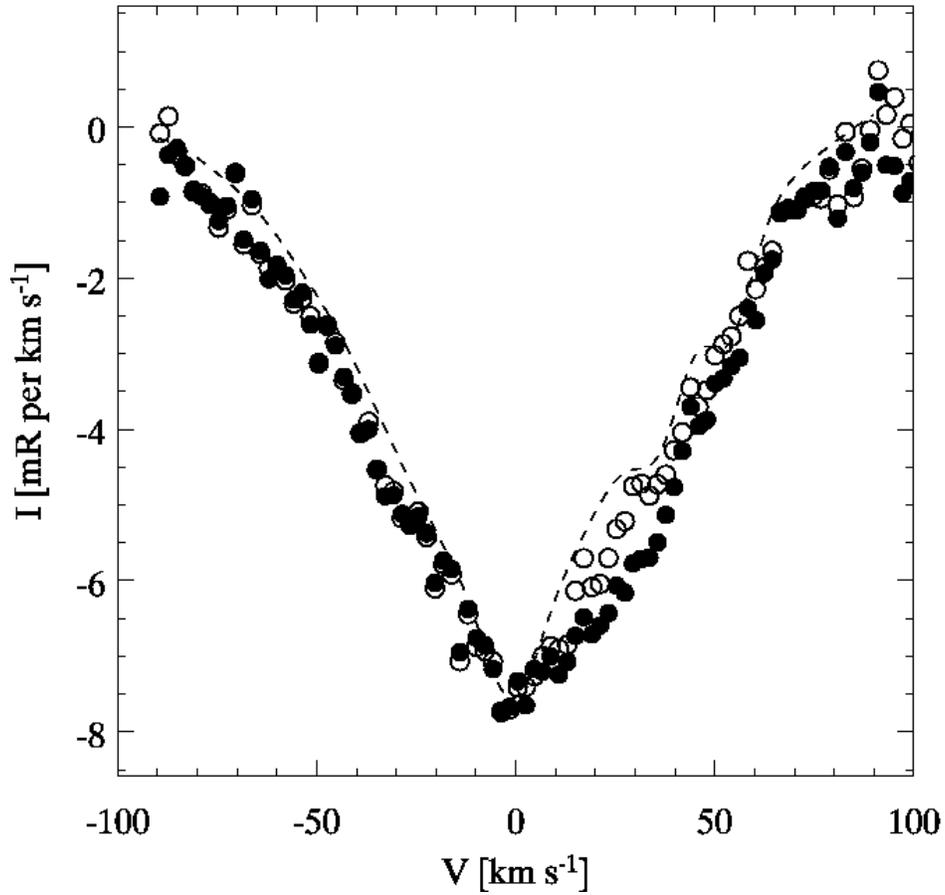}
\caption{Mean spectrum for $\epsilon \approx 174^{\circ}
- 188^{\circ}$, showing that the antisolar zodiacal profile is broadened
relative to the unperturbed solar profile (convolved direct solar 
spectrum; dashed curve).  The filled and open
symbols represent the profile with maximal and minimal corrections for the
atmospheric contamination, respectively (see \S3.1 and Fig. 2c). The 
``continua'' and minima of the profiles were made to coincide.
}
\end{figure}

\clearpage

\begin{figure}[t]
\epsscale{0.75}
\plotone{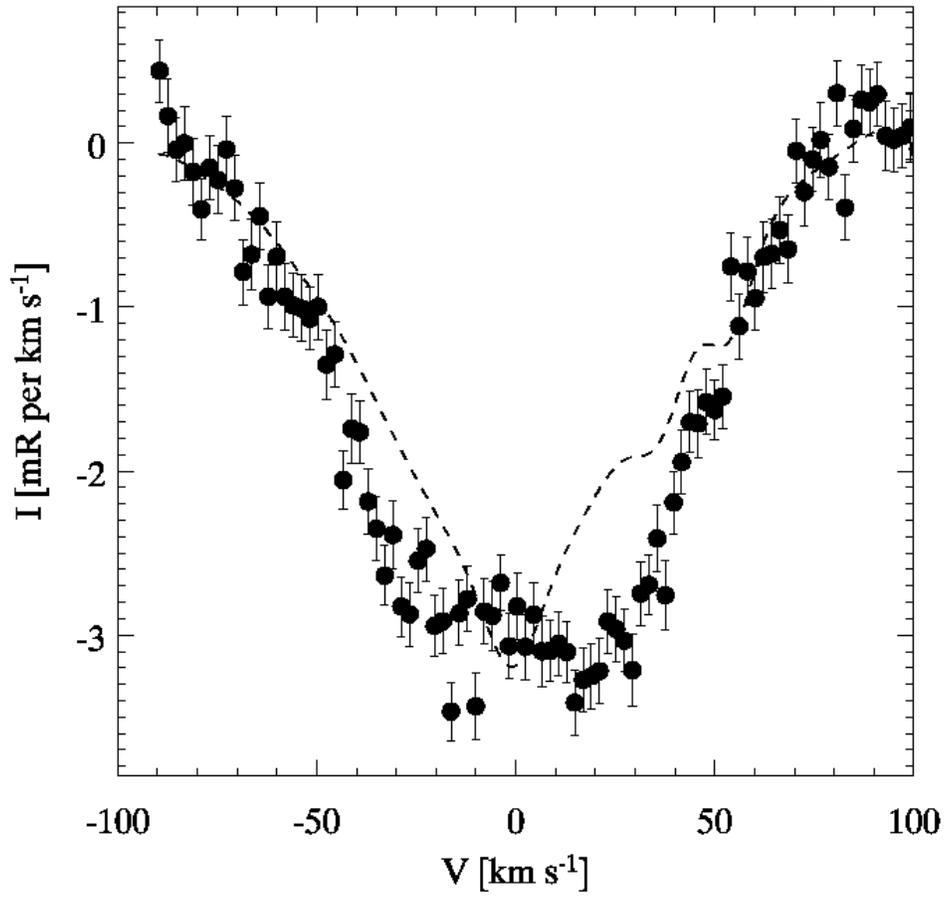}
\caption{Zodiacal spectrum toward the north ecliptic pole (NEP).  The 
unperturbed solar profile is included for comparison. The error bars 
represent the $\pm 1 \sigma$ Poisson noise.
}
\end{figure}

\clearpage

\begin{figure}[t]
\epsscale{0.75}
\plotone{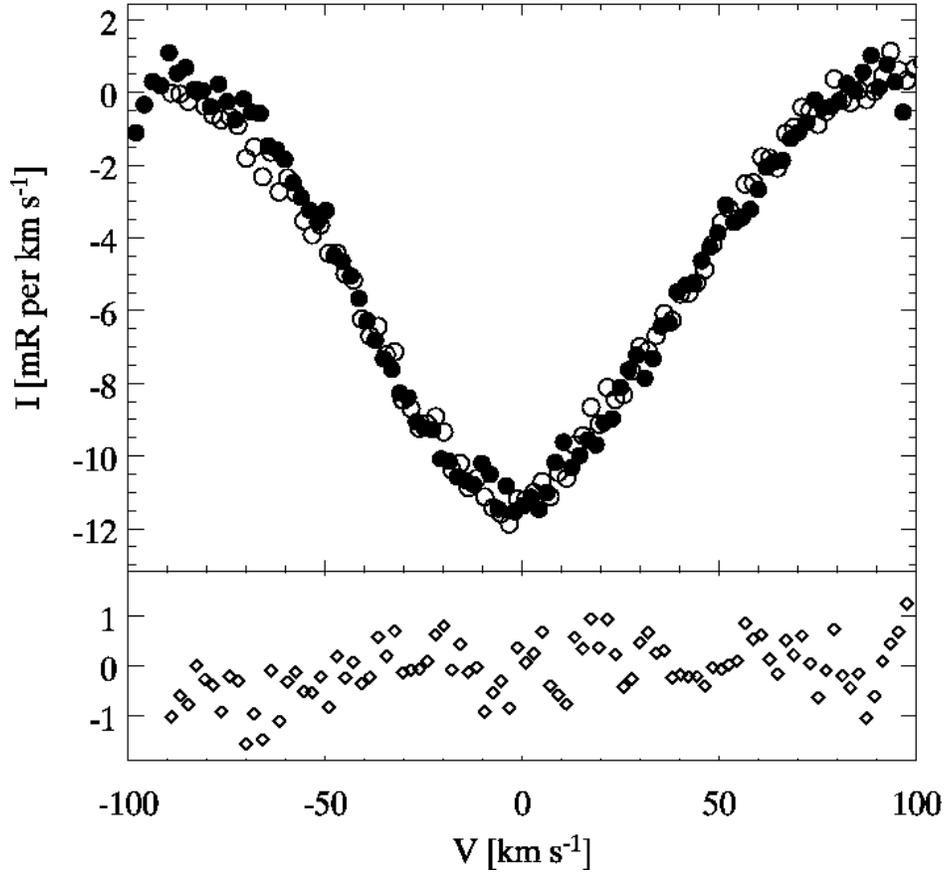}
\caption{Average spectrum between $\epsilon \approx 50^{\circ}$ and
70$^{\circ}$ (open symbols) and the average spectrum between $\epsilon
\approx 290^{\circ}$ and 310$^{\circ}$ (filled symbols) have been
superposed to search for asymmetries in the line profiles (see \S4.3).
To make the comparison easier, the former spectrum was shifted $-9.0$
km s$^{-1}$ with respect to the latter spectrum. The difference
between the two profiles is also plotted (open diamonds).
}
\end{figure}

\clearpage

\begin{deluxetable}{rrrrrrrrrrrrr}
  \rotate
\tablecaption{Summary of Observations and Results}
\tabletypesize{\small}
\tablewidth{0pt}
\tablenum{1}
\tablehead{ \colhead{$\epsilon$} & \colhead{$\beta$} & \colhead{Time} &
  \colhead{V}  & \colhead{$\Delta$V$_{min}$\tablenotemark{a}} &
  \colhead{$\Delta$V$_{max}$\tablenotemark{a}} &
  \colhead{W}  & \colhead{$\Delta$W$_{min}$\tablenotemark{a}} &
  \colhead{$\Delta$W$_{max}$\tablenotemark{a}} &
  \colhead{A}  & \colhead{$\Delta$A$_{min}$\tablenotemark{a}} &
  \colhead{$\Delta$A$_{max}$\tablenotemark{a}} \\
  \colhead{($^\circ$)} & \colhead{($^\circ$)} & 
  \colhead{(sec)} & \colhead{(km s$^{-1}$)} & \colhead{(km s$^{-1}$)} & \colhead{(km s$^{-1}$)}
   & \colhead{(km s$^{-1}$)} & \colhead{(km s$^{-1}$)} & \colhead{(km s$^{-1}$)} &
     \colhead{(R)} & \colhead{(R)}  & \colhead{(R)} }
\startdata
  47.5 &   0.0 &  240 &   -3.4 &  -0.5 &   +0.5 &   80.3 &  -1.4 &   +1.3 &   -1.67 & -0.04 &  +0.05 &  \\
  57.5 &   0.0 &  120 &    +2.1 &  -0.6 &   +0.7 &   76.1 &  -0.3 &   +0.5 &   -1.27 & -0.01 &  +0.03 &  \\
  57.9 &   0.0 &  600 &    +2.8 &  -0.6 &   +0.6 &   74.1 &  -0.2 &   +0.2 &   -1.31 & -0.01 &  +0.01 &  \\
  67.5 &   0.0 &  240 &    +5.7 &  -0.6 &   +0.5 &   78.1 &  -1.3 &   +1.3 &   -1.24 & -0.04 &  +0.04 &  \\
  67.9 &   0.0 &  600 &    +6.7 &  -0.8 &   +0.8 &   75.6 &   +0.3 &  -0.4 &   -0.98 &  0.00 &  0.00 &  \\
  72.5 &   0.0 &  120 &   +10.9 &  -0.8 &   +0.8 &   78.2 &   +1.3 &  -1.4 &   -0.95 &  0.01 & -0.01 &  \\
  77.5 &   0.0 &  120 &   +10.3 &  -0.8 &   +0.8 &   80.1 &  -1.6 &   +1.5 &   -0.78 & -0.04 &  +0.04 &  \\
  77.9 &   0.0 & 1800 &    +9.1 &  -1.0 &   +1.0 &   74.4 &   +0.7 &  -0.9 &   -0.77 &  0.00 &  0.00 &  \\
  82.5 &   0.0 &  120 &    +9.5 &  -1.1 &   +1.1 &   82.1 &   +2.0 &  -2.1 &   -0.79 &  0.01 & -0.01 &  \\
  87.5 &   0.0 &  120 &   +12.2 &  -1.2 &   +1.1 &   79.1 &   +1.9 &  -2.1 &   -0.70 &  0.01 & -0.01 &  \\
  87.9 &   0.0 &  600 &   +12.1 &  -0.8 &   +0.7 &   74.8 &  -1.7 &   +1.6 &   -0.68 & -0.04 &  +0.04 &  \\
  92.5 &   0.0 &  120 &   +14.1 &  -1.1 &   +1.1 &   81.8 &   +2.5 &  -2.6 &   -0.73 &  0.01 & -0.01 &  \\
  97.8 &   0.0 &  600 &   +11.6 &  -0.7 &   +1.2 &   73.1 &  -1.6 &   +2.4 &   -0.57 & -0.04 &  +0.05 &  \\
 107.8 &   0.0 &  600 &   +12.8 &  -1.4 &   +1.3 &   69.7 &   +1.9 &  -2.2 &   -0.49 &  0.00 &  0.00 &  \\
 127.9 &   0.1 &  600 &   +11.7 &  -1.5 &   +1.4 &   74.5 &   +2.3 &  -2.5 &   -0.50 &  0.00 & -0.01 &  \\
 137.5 &   0.0 &  120 &    +6.5 &  -1.6 &   +1.6 &   87.2 &   +2.6 &  -3.0 &   -0.67 &  0.02 & -0.02 &  \\
 137.8 &   0.0 &  600 &   +10.9 &  -1.1 &   +1.0 &   72.9 &  -2.4 &   +2.2 &   -0.49 & -0.04 &  +0.04 &  \\
 147.8 &   0.0 &  600 &   +10.5 &  -1.3 &   +1.3 &   69.8 &   +1.3 &  -1.6 &   -0.52 &  0.00 &   0.00 &  \\
 157.8 &   0.0 &  600 &    +6.9 &  -1.2 &   +1.1 &   72.8 &  -3.1 &   +2.8 &   -0.52 & -0.04 &  +0.04 &  \\
 167.8 &   0.0 &  600 &    +2.8 &  -1.2 &   +1.2 &   70.1 &  -3.4 &   +3.0 &   -0.47 & -0.04 &  +0.04 &  \\
 169.2 &   0.0 &  120 &    +1.4 &  -1.3 &   +0.8 &   78.7 &  -3.7 &   +2.8 &   -0.57 & -0.05 &  +0.04 &  \\
 174.2 &   0.0 &  120 &   -0.3 &  -1.1 &   +1.0 &   83.9 &  -3.0 &   +2.8 &   -0.66 & -0.04 &  +0.05 &  \\
 177.8 &   0.0 &  600 &    +0.9 &  -1.0 &   +1.3 &   75.4 &  -3.0 &   +3.3 &   -0.56 & -0.04 &  +0.05 &  \\
 179.3 &   0.0 &  120 &   -1.3 &  -1.0 &   +1.0 &   65.6 &  -3.6 &   +3.4 &   -0.53 & -0.04 &  +0.04 &  \\
 184.2 &   0.0 &  120 &   -1.6 &  -1.2 &   +1.0 &   71.2 &  -3.6 &   +3.1 &   -0.52 & -0.04 &  +0.04 &  \\
 187.8 &   0.0 &  600 &   -0.8 &  -1.2 &   +1.2 &   77.2 &  -3.9 &   +3.6 &   -0.54 & -0.04 &  +0.04 &  \\
 189.2 &   0.0 &  120 &   -4.1 &  -1.1 &   +1.1 &   84.5 &  -3.4 &   +3.1 &   -0.62 & -0.04 &  +0.04 &  \\
 197.8 &   0.0 &  600 &   -0.4 &  -1.2 &   +1.1 &   73.4 &  -3.7 &   +3.5 &   -0.51 & -0.04 &  +0.04 &  \\
 207.8 &   0.0 &  600 &   -3.9 &  -1.3 &   +1.3 &   75.2 &  -4.3 &   +3.9 &   -0.50 & -0.04 &  +0.04 &  \\
 239.0 &   2.0 &  300 &   -8.2 &  -1.2 &   +1.2 &   75.4 &  -4.2 &   +4.0 &   -0.53 & -0.04 &  +0.04 &  \\
 244.0 &   2.0 &  300 &  -11.0 &  -1.2 &   +1.3 &   80.9 &  -1.6 &   +1.5 &   -0.59 & -0.01 &  +0.01 &  \\
 249.0 &   2.0 &  300 &  -10.7 &  -1.1 &   +1.2 &   81.4 &  -4.7 &   +4.5 &   -0.57 & -0.04 &  +0.05 &  \\
 252.1 &   0.0 &  600 &  -10.9 &  -1.1 &   +1.1 &   80.4 &  -4.6 &   +4.3 &   -0.59 & -0.04 &  +0.05 &  \\
 254.0 &   2.0 &  300 &  -10.7 &  -0.9 &   +1.1 &   78.4 &  -5.0 &   +4.5 &   -0.57 & -0.04 &  +0.04 &  \\
 262.1 &   0.0 & 1800 &  -10.2 &  -0.9 &   +1.2 &   78.8 &  -1.2 &   -3.2 &   -0.74 & -0.01 & -0.06 &  \\
 264.0 &   2.0 &  300 &  -12.8 &  -0.6 &   +1.6 &   77.3 &  -3.9 &   +5.2 &   -0.57 & -0.04 &  +0.05 &  \\
 282.0 &   0.0 &  600 &  -10.7 &  -0.8 &   +0.8 &   77.7 &  -3.2 &   +3.2 &   -0.77 & -0.04 &  +0.04 &  \\
 283.9 &   2.0 &  300 &   -8.2 &  -0.9 &   +0.9 &   73.2 &  -3.5 &   +3.3 &   -0.67 & -0.04 &  +0.04 &  \\
 288.9 &   2.0 &  300 &   -7.7 &  -0.8 &   +0.8 &   77.2 &  -3.0 &   +2.9 &   -0.82 & -0.04 &  +0.04 &  \\
 293.9 &   2.0 &  300 &   -7.0 &  -0.8 &   +0.8 &   76.8 &  -2.8 &   +2.7 &   -0.84 & -0.04 &  +0.04 &  \\
 298.9 &   2.0 &  300 &   -5.4 &  -0.7 &   +0.7 &   74.0 &  -2.6 &   +2.4 &   -0.90 & -0.04 &  +0.04 &  \\
 303.9 &   2.0 &  300 &   -2.8 &  -0.6 &   +0.7 &   77.1 &  -0.5 &   +0.4 &   -1.15 & -0.01 &  +0.01 &  \\
 308.9 &   2.0 &  300 &   -2.0 &  -0.6 &   +0.6 &   77.2 &  -0.4 &   +0.3 &   -1.32 & -0.01 &  +0.01 &  \\
 312.0 &   0.0 &  600 &   -1.0 &  -0.4 &   +0.4 &   74.9 &  -0.3 &   +0.3 &   -1.94 & -0.01 &  +0.01 &  \\
 313.9 &   2.0 &  300 &    +0.4 &  -0.5 &   +0.5 &   74.8 &  -0.3 &   +0.2 &   -1.58 & -0.01 &  +0.01 &  \\
 318.9 &   2.0 &  300 &    +2.2 &  -0.4 &   +0.4 &   73.6 &  -0.2 &   +0.1 &   -1.84 & -0.01 &  +0.01 &  \\
 323.8 &   2.0 &  300 &    +5.1 &  -0.3 &   +0.3 &   77.0 &   +0.1 &  -0.1 &   -2.45 & -0.00 &  0.00 &  \\
 328.9 &   2.0 &  300 &    +7.6 &  -0.2 &   +0.2 &   75.2 &   +0.1 &  -0.1 &   -3.07 & -0.00 &  +0.01 &  \\
 333.9 &   2.0 &  300 &    +8.5 &  -0.2 &   +0.2 &   76.1 &   +0.1 &  -0.1 &   -5.06 &  0.00 &  0.00 &  \\
   NEP &  90.0 & 2580 &   -2.2 &  -2.5 &   +2.5 &   83.2 &   +0.4 &  -0.4 &  -0.36 &  0.00 &  +0.01   \\
  272  &  47   & 1800 &   -3.2 &  -2.1 &   +2.1 &   77.3 &   +1.4 &  -1.4 &   -0.39 &  +0.01 & -0.01   \\
 Twil  &  ight &  ... &    0.0 &   ... &   ... &   58.2 &   ... &   ... &   ...  &   ... &   ...   \\
 \enddata
 \tablenotetext{a}{$\Delta$X$_{min}$ ($\Delta$X$_{max}$) is the change in the corresponding value if
   the minimal (maximal) correction is used for the atmospheric lines,
   i.e. $\Delta$X$_{min} = $X$_{min}$ - X (see \S3)}
\end{deluxetable}

\end{document}